%% file: sample-acmsmall.tex
\documentclass[biblatex]{article}

\usepackage{graphicx}
\usepackage{multirow}
\RequirePackage[
backend=bibtex,
style=numeric
  ]{biblatex}

\usepackage{tikz}
\usetikzlibrary{arrows,decorations.pathmorphing,backgrounds,fit,positioning, decorations.pathreplacing, calc,shapes}
\usepackage{pgfmath}	

\usepackage{rotating}		
\usepackage{array}		
\usepackage{graphicx}	        
\usepackage{float}		
\usepackage{mdwlist}            
\usepackage{setspace}           
\usepackage{listings}		
\usepackage{bytefield}          
\usepackage{tabularx}		
\usepackage{multirow}	        
\usepackage{amsmath}
\usepackage{subcaption}
\usetikzlibrary{matrix}
\usepackage{pgfplots}
\usepackage{cancel}
\usepackage{afterpage}
\usepackage{longtable}
\usepackage{textpos}

\usepackage{url}                
\usepackage{hyperref}
\usepackage[all]{hypcap}

\newcommand{\repeatthanks}{\textsuperscript{\thefootnote}}

\begin{document}

\title{Determining the Optimal Frequencies for a Duplicated Randomized Clock SCA Countermeasure}

\date{}
\author{Gabriel Klasson Landin\thanks{Both authors contributed equally to this manuscript.} \and
Truls Jilborg\repeatthanks   \and\\Department of Electrical Engineering, \\
Royal Institute of Technology (KTH)\\
Electrum 229, 196 40 Stockholm, Sweden \\
\{gablan, trulsj\}@kth.se}



\maketitle  
\begin{abstract}
Side-channel attacks pose significant challenges to the security of embedded systems, often allowing attackers to circumvent encryption algorithms in minutes compared to the trillions of years required for brute-force attacks. To mitigate these vulnerabilities, various countermeasures have been developed. This study focuses on two specific countermeasures: randomization of the encryption algorithm’s clock and the incorporation of a dummy core to disguise power traces. 

The objective of this research is to identify the optimal frequencies that yield the highest level of randomness when these two countermeasures are combined. By investigating the interplay between clock randomization and the presence of dummy cores, we aim to enhance the overall security of embedded systems. The insights gained from this study will contribute to the development of more robust countermeasures against side-channel attacks, bolstering the protection of sensitive information and systems. 

To achieve this, we conduct simulations and perform side-channel attacks on an FPGA to establish the relationship between frequencies and the resulting protection. We break the encryption on a non-duplicated circuit and note the least amount of measured power traces necessary and the timing overhead. We do this for all sets of frequencies considered which gives a good indication of which sets of frequencies give good protection. By comparing the frequencies generated with those from the duplicated circuit we use similar conclusions to prove whether a frequency set is secure or not. 

Based on our results we argue that having one frequency lower than half of the base frequency and the other frequencies being close but not higher than the base gives the highest security compared to the timing overhead measured.
\end{abstract}




\paragraph{keywords:}
Side channel attack, Cryptography, Correlation power analysis, Chipwhisperer, Frequency, Hardware security


\maketitle

\section{Introduction}
\label{ch:introduction}

Breaking AES encryption through a brute force attack would take over  $10^{18}$ years~\cite{AES_Bruteforce}. The encryption typically uses a 128-bit key, which means that the sets of possible keys are $2^{128}$ but it can also use a 192 or 256-bit key for even more possible keys. Consequently, brute force attacks, which involve exhaustively trying all possible keys, are computationally infeasible and practically impossible.
Because of this, another approach has been researched: the possibility of a divide-and-conquer method. This method would break down the key into smaller chunks and attack each chunk. Instead of guessing $2^{128}$ different keys, one would only need to guess $2^8$(256) keys. This can be done by exploiting vulnerabilities in the physical implementation of the encryption algorithm and measuring the power consumption of the chip performing the encryption. It is possible to correlate the power consumed with the part of the key being used. This correlation can then be used to check if the guessed key is correct, and this type of attack is called a side-channel attack, or SCA for short.

For integrated circuits, particularly Field-Programmable Gate Arrays \linebreak (FPGAs), side-channel attacks present a significant challenge. These circuits often utilize hard-coded keys that are difficult to change once set. If a perpetrator manages to perform a side-channel attack and get the key, this would lead to all future encryptions being compromised, jeopardizing the confidentiality and integrity of sensitive data.

To counteract these threats, this research project aims to develop strategies to mitigate side-channel attacks and make power traces harder to correlate with specific encryption keys. This thesis will be based on research where AES is performed with random frequencies~\cite{Martin} and when a dummy AES core is introduced~\cite{Duplication}. By minimizing the leakage of information through careful selection and optimization of frequencies, the goal is to reduce the effectiveness of side-channel attacks and enhance the overall security of AES implementations on FPGAs.

\section{Our contributions}

We have established some of the relational behaviour of frequencies used for the randomized clocks in~\cite{Martin} and~\cite{Duplication} and have come to a conclusion about how to set the frequencies depending on the requirement of the system. Frequencies that give a high resistance against CPA attacks are given, and suggestions of frequencies that theoretically should provide good countermeasures against DL/MLP.

\section{Background}
\label{sec:background}

\subsection{Previous work}
\label{subsec:previous work}
Side-channel attacks pose a significant threat to cryptographic systems. Paul Kocher et al.~\cite{DPA_RSA} showed that it is possible to measure the power consumed when a circuit performs an encryption using the Data Encryption Standard (DES). In these traces, it is possible to distinguish the different operations of the encryption algorithm. The same can be found when measuring the power consumption of Advanced Encryption Standard (AES~\cite{NIST-AES})~\cite{DPA_Intro, DPA&CPA_AES}.
Using these power traces it is possible to perform Differential Power Analysis (DPA), Correlation Power Analysis (CPA) and other power analysis attacks. 

Ever since SCAs were first discovered, it has been a constant race between finding new ways to exploit side-channel leakage and to protect against it. Researchers have developed a number of countermeasures to protect against these attacks. These countermeasures include masking, shuffling, hiding, and power analysis-resistant logic.

Masking hides sensitive values by splitting them and performing operations (often XOR) on each part with a random mask. This reduces leakage and makes correlating power consumption to the key more difficult. In the end, the operation is reversed to get the correct output~~\cite{Masking}. Shuffling rearranges the order of instructions to prevent timing attacks~~\cite{shuffle}, and hiding involves concealing the input or output of a cryptographic operation to prevent power analysis attacks~\cite{hiding}. Power analysis-resistant logic adds additional logic to the design to make it harder for attackers to extract information from the system~\cite{resistant-logic}.

While these countermeasures can be effective against many types of side-channel attacks, they are not always foolproof. For example, higher-order attacks can still compromise masking~\cite{cryptoeprint:2022/1713}, and shuffling may be ineffective if the implementation is leaky~\cite{shuffle}.

Clock randomization is a promising countermeasure for protecting a cryptographic key from side-channel attacks. One such countermeasure, RFTC (Random Frequency Tuning Countermeasure), was developed by~\cite{RFTC} RFTC uses dynamic frequency scaling to randomly adjust the clock signal's frequency during runtime, making it harder for attackers to correlate power consumption with the cryptographic key. RFTC is implemented on an FPGA using dynamic reconfiguration, which makes it easy to test and implement in hardware.

However, researchers~\cite{Martin} found that attacking the first round of AES and oversampling the signal at a rate significantly higher than the Nyquist rate~\cite{DSPfirst} of the signal can still extract the key with enough traces. To overcome this issue,~\cite{Duplication} used a dummy AES core with a different random frequency and key but the same plaintext. They ran two duplicate encryption cores, each controlled by an unstable clock, creating a combinatorial problem when synchronizing. This approach requires a brute-force attack with \(2^n\) attempts, where n is the number of traces required for a successful attack.

The results of~\cite{Duplication} were promising, with neither MLP (Multilayer Perceptron) nor FFT (Fast Fourier Transform)  able to extract the key from 10 million traces when using two randomized clocks and duplication. However, it is still unclear what effect the possible values of the two frequencies can have.

\subsection{Theory}
\label{subsec:theory}

AES is a popular encryption algorithm and is recognized as the standard by the US government~\cite{NIST-AES}. It performs the encryption in a number of rounds with a key  that is expanded into subkeys that are used for each round. The key size is proportional to the number of rounds - 10/12/14 rounds for a key size of 128/192/256 bit - in order for the key expansion to create homogeneous subkeys for each round. In side-channel attacks, the last and the first rounds can be more easily attacked since they use a known input and a known output in the encryption algorithm for that round~\cite{leakage}. When attacking the first round a MLP attack is required and a regular CPA attack against FPGA boards usually target the last round in order to be successful~\cite{Martin}.   

\begin{figure}
  \begin{subfigure}{0.5\textwidth}
    
    \includegraphics[width=\linewidth]{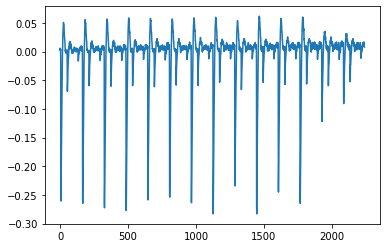}
   
    \caption{\centering The first/last round of AES is easy to find with a non randomized clock.}
  \end{subfigure}%
  \begin{subfigure}{0.5\textwidth}
    
    \includegraphics[width=\linewidth]{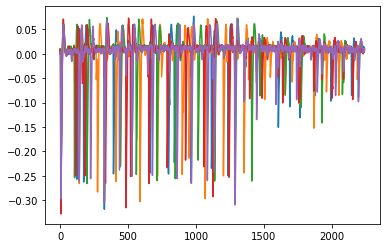}
    \caption{\centering 10 Captured power traces with randomized clock.}
  \end{subfigure}
  \caption{\centering Comparison of traces in a non randomized implementation and a randomized implementation.}
  \label{figure:randomVSnonerandom}
\end{figure}

CPA is the attack method used in this paper and consists of a few steps~\cite{CPA, NAE_CPA}. 
A model for the power consumption is selected, most commonly Hamming weight or Hamming distance. Hamming weight is calculated as the number of bits set to 1, which is assumed to correlate to the power consumed by the operation. Hamming Distance is the number of bits flipped, which corresponds to the distance in power consumption. The encryption key is divided into subkeys and each byte is attacked one by one. For each byte, all possible values for the byte are considered. Using the power model and input/output a power hypothesis is calculated, usually in the form of \(HW(key\oplus input/output)\). The Pearson correlation coefficient, \autoref{Pearsoncof}~\cite{Pearson} is used between the calculated power hypothesis for each key guess and the measured power from each trace.
\begin{equation}
 \frac{\sum\limits_{i=1}^{n} (X_i - \bar{X})(Y_i - \bar{Y})}{\sqrt{\sum\limits_{i=1}^{n} (X_i - \bar{X})^2 \sum\limits_{i=1}^{n} (Y_i - \bar{Y})^2}}
\label{Pearsoncof}
\end{equation}
\begin{center} \(n\) is the number of traces, \(X\) is the calculated power hypothesis for the key guess and \(Y\) is the measured power.
\end{center}
The key guess with the highest correlation will be the correct guess in a successful attack. For the CPA attack to be successful, a lot of traces have to be captured in order to get an accurate value for the correlation with the hamming distance, which can be hard without synchronization if the clock is randomized, see \autoref{figure:randomVSnonerandom}. 

Finding the optimal frequencies to improve the security against an SCA using AES with a randomized clock can be difficult and has a lot of factors to be taken into consideration. 

In~\cite{RFTC} they use the formula \(_{r+n-1}C_r\)\footnote{\(_{n}C_k\) is an alternative notation for \(\binom{n}{k}\)} - where n is the number of unique frequencies and r is the number of rounds in AES - to calculate the number of different completion times for AES. They use it as a measure for security since it is equal to the number of non synchronized last rounds for each trace. This indicates that the system is more secure the higher the number of unique frequencies that are in use. Using a setup like the one in~\cite{Martin} with 4 frequencies connected to a MUX controlled asynchronous by a base frequency randomly switching between the frequencies, the optimal choice for the frequencies would be the ones yielding the most unique frequencies. By choosing frequencies that are relative primes to each other, it will maximize the coincidence period and thereby potentially create more unique frequencies. It is a bit unclear whether this setup supports continuous values for the base frequencies, but having the GCD between them approach the limit of zero would create an infinitely long coincidence period for potentially an unlimited number of unique frequencies. This, however, raises the question of the sensitivity of reading the rising edge. If the difference  between two periods is too small, it might not be recognized. It could also be read as minimal differences or noise causing more randomness.\\

Another thing that the formula used in~\cite{RFTC} does not take into consideration is the probability of a unique frequency appearing, having thousands of unique frequencies will not cause much randomness if it is only one frequency that appears the majority of the time. In order to see the results from simulations and tests, a fast Fourier transform (FFT)~\cite{DSPfirst} will be made for each sample of frequencies. This will convert the traces from a time spectrum to a frequency spectrum that shows the number of different frequencies.

Calculation of relations between clock signals with high frequencies is difficult, so we will use a uniform distribution and a version of binomial distributions~\cite{Sannolikhetsteori} in order to predict the behaviour of frequencies.  

\section{Problem}

It is crucial to find the optimal frequencies, since the effectiveness of the countermeasure developed by~\cite{Duplication} is heavily affected by it. The study conducted by~\cite{RFTC} had selected frequencies that avoided overlapping, which is to be expected when in a circuit comprising one AES core. 
Dealing with two asynchronous cores however takes this problem to a new dimension. Apart from finding the optimal frequency to avoid overlap in one of the clocks, there will be two separate clocks that might overlap with each other. It is also unknown how this overlap will affect the security. As~\cite{Duplication} mentions it could be a good thing since the overlap is varying in time and therefore introduce additional randomness to the circuit.~\cite{Duplication} also mentions that this varying overlap will have a positive effect on security against methods to increase SNR, e.g. by repeating and averaging.

Moreover, it is essential to understand the impact of frequencies on the circuit's performance, as it affects the overall effectiveness of the\linebreak countermeasure. The study will enable researchers to identify the effect of frequencies on the circuit, thereby helping them to make informed decisions when designing and implementing the countermeasure.

\section{Goals}

Finding the optimal frequencies is a critical step in developing an effective countermeasure. The analysis of previous research and the impact of\linebreak frequencies on the circuit's performance will assist in developing an efficient solution that can provide maximum protection against attacks. Our goal is therefore to find these optimal frequencies as well as the reason why they are optimal.

\section{Method}
\label{ch:methods}

This section describes how we approach the problem, starting off by stating what we will do. It continues by describing the theoretical process and explains some of the findings. Finally, the practical research process is described, where the equipment is listed, and the planned experiments are explained.

\section{Method introduction}
\label{sec:methodintroduction}

We will create an efficient method to calculate the permutations of 4 frequencies randomly chosen by an asynchronous clock to acquire an exact probability for each frequency that appears during a coincidence period. This will make the old formula more realistic for evaluation of security. Through the utilization of the algorithm, we hope to identify connections and evidence regarding the behaviour of frequencies in relation to each other. We will use the results in order to maximize randomness and to find a set of frequencies with a uniform distribution. Just by examining the frequencies it can be seen that each base clock cycle will create one out of 4x4 permutations with some exceptions:

 \begin{enumerate}
     \item Frequencies slower than the base frequency will occasionally not have a rising edge relative to the base frequency. The created frequency will then stretch through more than one base cycle creating another set of \(4*4\) permutations, which will increase the set of unique frequencies, see \autoref{figure:frequencies}.
     \item Higher frequencies on the other hand will occasionally reach two rising edges during its duty time, creating two frequencies instead, one of them being the original frequency chosen. This means that frequencies faster than the base frequency will appear more frequent relative to how much faster they are and the system might lose randomness, see \autoref{figure:frequencies}.
     \item Another exception is when the active frequency is low when the rising edge of the base clock change the active frequency to a frequency that is high at the time of the switch. This too will create an extra frequency during the  base period with the rising edge same as the base clock, see \autoref{figure:frequencies}.
 \end{enumerate}

\input{Figures/frequencyfigure} 

By taking the things in \autoref{sec:methodintroduction} into consideration, we intend to find 4\linebreak frequencies in relation to a base clock that generates a uniform set of frequencies with as many unique frequencies as possible. 

\section{Method process}
\label{sec:methodprocess}

In order to find unique frequencies, the current and the previous period of the base clock will be examined. What is interesting here are the exceptions to the 4x4 permutations that can happen. \autoref{figure:Perm_one_missing_rising_edge} shows the first exception when one frequency does not have a rising edge and how it increases the permutations from \(4*4\) to \(4*(3+4)-4 \)\footnote{Minus the repeating permutations in the next clock cycle in order to avoid duplication} . In \autoref{tab:Permutations} it is visible that the number of frequencies having no rising edge in the current period affects the permutations differently and that only one of the 4 frequencies should be without a rising edge for the maximum number of permutations. 

\input{Figures/Permutation_table}

\begin{figure}[H]
  \begin{subfigure}{0.5\textwidth}
    \centering
    \input{Figures/Perm_without_rising_edge}
    \caption{\centering The doted lines show the extra permutations created.}
  \end{subfigure}%
  \begin{subfigure}{0.5\textwidth}
    \centering
    \input{Figures/Perm_after_rising_edge}
    \caption{\centering The doted lines are already accounted for in the next cycle.}
  \end{subfigure}
  \centering
  \caption{Permutations with one missing rising edge.}
  \label{figure:Perm_one_missing_rising_edge}
\end{figure}
However, assuming that there is only one frequency that is missing a rising edge, the extra 12 permutations will only be in effect $\frac{1}{4}$ of the time giving each of the extra frequencies a probability of $\frac{1}{64}$ instead of $\frac{1}{16}$. This is not desirable, for a uniform set of frequencies they all should appear with equal probability. Having two rising edges will even the probability, but will not create more permutations. This indicates that the probability of having a lot of unique frequencies is a trade-off towards having a uniform distribution.
In reality, the probability of a rising edge during a base cycle is not evenly distributed and can be calculated as a uniform distribution~\cite{Sannolikhetsteori}.

\begin{equation}
  F(x) =
  \begin{cases}
    \frac{T_{b}}{T_{i}} & \text{$T_{i}$}\geq \text{$T_{b}$} \\
    1 & \text{$T_{b}$}> \text{$T_{i}$}
  \end{cases}
  \label{uniformdistribution1}
\end{equation}

\autoref{uniformdistribution1} shows how the probability of a rising edge can be calculated, where $T_{i}$ is the period of the frequency to be examined and $T_{b}$ is the period of the base clock. 

Knowing the probability of having a rising edge in the period, it is possible to calculate each individual set of permutations and cumulatively add them together. This is done with a version of binomial distribution~\cite{Sannolikhetsteori} and will give the probability of rising edges occurring at the same time. 

\[
    \sum P_i\cdot Q_i
\]
Where  $P_i$ holds the  number of permutations that can generate the wanted number of rising edges and $Q_i$ holds the complement of the permutations. For example, having 4 frequencies with the probabilities  $p_1$, $p_2$, $p_3$ and $p_4$ of having a rising edge during the base period will generate permutations based on the wanted number of rising edges that should occur at the same time. Adding them together gives \autoref{equation:perm_with_certain_number_of_RE}. 
\\
\input{Figures/Probability_no_rising_edge}

As can be seen in \autoref{tab:Permutations} you want at least 2 rising edges at the same time in order to avoid the loss of permutations. Frequencies higher than the base clock will always generate a rising edge. Therefore, having two faster frequencies will therefore fulfil this criterion. The other one or two frequencies, depending on what is wanted, should be set as low as possible. However, frequencies that deviate from the base frequency by more than a factor of \(0.5\) have a risk of occasionally lining up a low or high signal for the entire base cycle. This means that \autoref{tab:Permutations} has to be extended to include permutations two cycles ahead. As can be seen in \autoref{equation:perm_with_certain_number_of_RE}, new permutations have a lower probability to be in effect than the rest and the ones created two cycles ahead will have an even lower probability that will cause our distribution to be less uniform. Having two of these slower frequencies will also cause a risk of them having the same value after each other. Going from a full base cycle of a low signal to another cycle of a low signal will cause a  period that is as long as this idles. This will cause a lot of time overhead for the implementation, and a limit for the lowest frequency (at least more than one) should be set to a factor of \(0.5\) of the base frequency.
\\

The second exception will not affect the permutations, but it will add one of its fundamental frequencies with a likelihood described by a similar uniform distribution as the first exception:
\begin{equation}
  F(x) =
  \begin{cases}
    0 & \text{$T_{i}$}> \text{$T_{b}$} \\
    \frac{T_{b}-T_i}{T_{i}} & \text{$T_{b}$}\geq \text{$T_{i}$}
  \end{cases}
\end{equation}
While this increases the base cycle to have two frequencies instead of one, giving it a better time overhead, it should be noted that the fundamental frequencies already appear with a probability of $\frac{1}{4}$. In order to maintain a uniform distribution, frequencies higher than the base clock should be avoided. However, with exception 1 in mind it's necessary to have some higher frequencies, but these should be selected carefully. At a factor \(2\) of the base clock they will cross a breakpoint and have an extra period during each base cycle with a risk of having 2 extra periods.
\\ 

The third and last exception to the permutations happens when an active frequency that is low switches to a frequency that is high. This will create a new rising edge along the base clock's rising edge. This will separate the frequency that is created, and there will be two periods in the base cycle. However, if two frequencies in the upcoming cycle both are high, they will both generate the same frequency for a less uniform distribution.
Considering that all the fundamental frequencies have a duty cycle of 50\%, all four frequencies will have a 50/50 distribution of high and low signals, giving this exception an average probability of 1/4.  It is possible to manipulate the probability by manually selecting frequencies that overlap in a certain pattern. There is however to our knowledge no  way to do this for higher frequencies without using phase shifting. Without a good way to control this exception, there is not much to examine further. It should be noted that frequencies lower than the base clock has a risk of generating the base clock's frequency with this exception. This happens if there is no rising edge in the upcoming frequency and the exception occurs again. This will cause the  frequency of the base clock to appear relative to the probability of a rising edge, see \autoref{figure:Slow_Freqs_Base_Clock}. While this will cause a less uniform distribution, it will reduce some of the time overhead caused by slower frequencies described in exception 1.

\begin{figure}
\begin{tikzpicture}

    \def\base{pi/5}
    \def\fone{3.9}
    \def\ftwo{4.1}
    \def\fthree{13}
    \def\ffour{17}
    
  \begin{axis}[
    width=\textwidth,
    height=16cm,
    xlabel={Time},
    samples=500,
    domain=0:2*pi,
    ymin=-12,
    ymax=8,
    axis lines=none,
    xtick={0,\base,2*\base,3*\base,4*\base, 5*\base, 6*\base, 7*\base, 8*\base, 9*\base, 10*\base, 11*\base, 12*\base},
    xticklabels={$0$, 1, 2,3, 4, 5, 6, 7, 8, 9, 10, 11, 12},
    ytick=\empty,
    clip=false
  ]
  
    \addplot+[red,line width=0.8pt,mark=none,domain=0.5:2*pi, solid] {5 + sign(sin(deg((3.9*x)+(pi)))};
    
    \addplot+[blue,line width=0.8pt,mark=none,domain=0.5:2*pi, solid] {2 + sign(sin(deg(4.1*x)))};

     \addplot+[red,line width=0.8pt,mark=none,domain=0.5:\base, solid] {-1 + sign(sin(deg(\fone*x)))};
     \addplot+[blue,line width=0.8pt,mark=none,domain=\base + 0.01:2*\base, solid] {-1 + sign(sin(deg(\ftwo*x)))};
     \addplot+[red,line width=0.8pt,mark=none,domain=2*\base + 0.01:3* \base, solid] {-1 + sign(sin(deg((\fone*x)+pi)))};
     \addplot+[blue,line width=0.8pt,mark=none,domain=3*\base + 0.01:4* \base, solid] {-1 + sign(sin(deg(\ftwo*x)))};
     \addplot+[red,line width=0.8pt,mark=none,domain=4*\base + 0.01:5* \base, solid] {-1 + sign(sin(deg((\fone*x)+pi)))};
     \addplot+[red,line width=0.8pt,mark=none,domain=5*\base + 0.01:6* \base, solid] {-1 + sign(sin(deg((\fone*x)+pi)))};
     \addplot+[blue,line width=0.8pt,mark=none,domain=6*\base + 0.01:7* \base, solid] {-1 + sign(sin(deg(\ftwo*x)))};
     \addplot+[red,line width=0.8pt,mark=none,domain=7*\base + 0.01:8* \base, solid] {-1 + sign(sin(deg((\fone*x)+pi)))};
     \addplot+[blue,line width=0.8pt,mark=none,domain=8*\base + 0.01:9* \base, solid] {-1 + sign(sin(deg(\ftwo*x)))};
     \addplot+[red,line width=0.8pt,mark=none,domain=9*\base + 0.01:10* \base, solid] {-1 + sign(sin(deg((\fone*x)+pi)))};

    \pgfplotsinvokeforeach{1,2,...,12}{
        \addplot+[black,line width=0.8pt,mark=none,const plot, dashed] coordinates {(#1*pi/5, -2) (#1*pi/5,             7)};
            \node at (axis cs: #1*pi/5, -3) {$T_{#1}$};
     }

     \node at (axis cs: 0, 5) {F1:};
      \node at (axis cs: 0, 2) {F2:};
      \node at (axis cs: 0, -1) {Fout:};

  \end{axis}
\end{tikzpicture}
\caption{Example of how low frequencies will generate the base clock frequency}
\label{figure:Slow_Freqs_Base_Clock}
\end{figure}

\section{Research Process}
\label{sec:researchProcess}

\subsection{Test environment/test bed/model}

For simulations the following script is used~\cite{Simulations}.
The equipment used is the CW1173 ChipWhisperer-Lite~\cite{NAE-CW1173} and a CW305 Artix 7~\cite{NAE-CW305} FPGA target board. Using the Chipwhisperer library~\cite{NAE-git} we used the following capture script~\cite{Capture_Script} which uses the Chipwhisperer capture api~\cite{Capture_Api} for capturing the traces used. The traces were analyzed using the following script~\cite{Attack_Script} which uses the chipwhisperer analyzer api~\cite{Analyzer_Api} for finding the hamming distance between rounds 9 and 10 of AES encryption.

\subsection{Data validity}
\label{ch:DataValidity}

To validate the results obtained from the previous experiments and simulations, side-channel attacks were conducted on the aforementioned board. The purpose of these attacks was to assess the behaviour of the identified frequencies when executed on actual hardware.

Each set of frequencies was tested using varying numbers of traces, with a higher number of traces indicating better frequencies. The same plaintext and key were utilized for each test, and the traces were collected sequentially with no difference in environment or input. Different segments of the traces were examined, since certain traces exhibit more leakage than others. A total of 30,000 traces were gathered and tested for each frequency set.

The algorithm begins by removing all traces that are incorrectly measured, for example where the chipwhisperer has missed the encryption window. The remaining traces are then synchronized, placing the peak of each round in the same location as all other traces. This synchronization removes all traces where the oversampling is insufficient, as described in~\cite{Martin}. Because of limitations in the algorithm, it also removes traces where the peaks are too close together such that it is too hard to distinguish individual peaks. 

The algorithm iterates through different segments of the traces and attempts to attack each permutation. When a key is discovered, the window of traces is narrowed down, and the iteration begins again. This ensures that the minimum number of traces required will be identified, regardless of the location of the high-leakage traces. 

\section{Results and Analysis}
\label{ch:resultsAndAnalysis}

This section starts off by showcasing the results. The simulated frequencies are listed and the results from the experiments are described and explained before comparing the two. The second part is the analysis, where the results are explained in more detail before discussing their implications.

\section{Results}

\subsection{Theory results with simulations}

Based on the method described in \autoref{sec:methodprocess}, we wanted to test frequencies that are likely to give us the permutations from \autoref{tab:Permutations}  and analyse how they compare to each other. \autoref{tab:simulation_results} presents 7 sets of frequencies and each column represents the following:
\begin{enumerate}
    \item The frequency group, each group represents the relationship between the frequencies and the base frequency. Each set is selected based on the different permutations from \autoref{tab:Permutations}.
    \item The precise values of the frequencies used.
    \item A graph, which presents the frequencies generated displayed in a histogram. The x-axis represents the frequency, while the y-axis represents the number of occurrences.
    \item Total number of rising edges generated based on a sample size of 32000 base clock cycles
    \item Number of unique frequencies generated.
\end{enumerate}

\include{Figures/newSim}

In \autoref{tab:simulation_results} we can see that most of the frequencies generated will not be unique. This comes from the input frequencies being more dominant, as seen in exception 2 above. The same can be said about exception 3, where the base frequency will be generated during slow frequencies. Therefore, the graphs will show spikes on high frequencies or on the base frequency when few high frequencies are used. 

\subsection{Attack results}

The results of the validity tests are displayed in \autoref{tab:attack_results}. Each row is a set of frequencies where 30,000 power traces were gathered. Each column represents the following:
\begin{enumerate}
\item The frequency group, each group represents the relationship between the frequencies and the base frequency. Each set is selected based on the different permutations from \autoref{tab:Permutations}.
  \item Represents the least amount of traces required to get the key.
  \item The percentage of failed encryptions, where the output ciphertext of the CW305 was incorrect. A higher number of failed encryptions could mean that the frequencies used make the circuit unstable, which forces re-encryptions and a high timing overhead.
  \item Percentage of traces that were either not a power trace or the peaks were too close to each other. This is also where traces with too little oversampling are discarded.
  \item The index of the last peak of an encryption. Represents the maximum amount of time required for an encryption using the provided frequencies.
  \item The worst timing overhead of the selected frequencies. Represented as the longest time elapsed for the clock to reach the last round compared to a non-randomized clock using the base clock. 
  \item The mean value of time overhead compared to the time taken without randomization using the base clock. These values are based on simulation.
\end{enumerate}
\input{Figures/Trace_table}

As can be seen in \autoref{tab:attack_results} frequency types \hyperref[tab:simulation_results]{"Two low, one above half, one lower than half"}\hyperref[tab:attack_results]{(2)} and  \hyperref[tab:simulation_results]{"Three low, one lower than half"}\hyperref[tab:attack_results]{(6)} have the lowest calculated overhead leading to them being the most consistently fast sets of frequencies. Other sets might have some frequencies that are faster, but overall they are the fastest.
\hyperref[tab:simulation_results]{"Two low, one above half, one lower than half"\hyperref[tab:attack_results]{(2)}, "Two high, two lower than half"\hyperref[tab:attack_results]{(3)}, "Three high, one lower than half" and "Three low, one lower than half"\hyperref[tab:attack_results]{(6)}} all have a high amount of traces required to break the encryption, leading them to be more difficult to attack.
\hyperref[tab:simulation_results]{"Two low, one above half, one lower than half"\hyperref[tab:attack_results]{(2)}, "Three high, one above half"\hyperref[tab:attack_results]{(7)} and especially "Two high, one above half, one lower than half"\hyperref[tab:attack_results]{(4)}} all have a significantly higher amount of failed encryptions, leading to those sets of frequencies being more unstable.
\hyperref[tab:simulation_results]{"Two low, one above half, one lower than half"\hyperref[tab:attack_results]{(2)}, "Three high, one lower than half"\hyperref[tab:attack_results]{(5)} and "Three low, one lower than half"\hyperref[tab:attack_results]{(6)}} have a high amount of traces removed for the synchronization, which could indicate that they have a lower leakage with a higher oversampling factor being required when gathering traces. \\
The timing overhead is calculated by dividing the maximum time spent encrypting using randomization by the maximum time spent encrypting using no randomization. It therefore gives the maximum overhead of the given frequencies.
The simulated overhead on the other hand is calculated from the mean time spent, giving a mean overhead for the given frequencies.

\include{Figures/Relation_comparison}

\autoref{tab:relation_comparison} shows that minor changes to the input frequencies have little to no effect. It is not until their relation to the base frequency is altered that there is an effect. The exception is when one of the frequencies is higher but comes too close to the base frequency, it leads to more encryption errors.

\subsection{FFT results of single and multicore AES}
\label{sub:fft_results}
\autoref{fig:FFT_single} and \autoref{fig:FFT_double} shows the correlation between simulated frequencies and actual frequencies to strengthen the results from the theoretical part. In \autoref{fig:FFT_single} at first glance it looks like the graphs of FFT and Simulations look quite different. The peaks are not in the same locations and are often not equally high. It is however possible to see similarities in the shapes and amount of peaks present. When plotting the graphs, a value called "bin" is used, which represents how values should be grouped into bars. We assume that this and the fact that the FPGA is not able to select the exact frequencies entered to be the reason for the mismatch. The Simulation also scales down the frequencies in order to not use as much CPU and memory, meaning that the simulation and testing uses different frequencies. The frequencies relate to each other in the same way but are not necessarily the same. This could explain the difference in height of the y axe's largest peaks compared to the rest, faster frequencies generates more unique frequencies compared to the dominant one, section: \ref{ch:methods}. As can be seen in \autoref{fig:FFT_double} the last two comparisons differ a lot in magnitude but have the same shape, which is to be expected.

\subsubsection{Single core vs simulations}
Each graph in \autoref{fig:FFT_single} symbolizes the Fast Fourier Transform of the powertraces of a set of frequencies. As mentioned in \autoref{sub:fft_results}, there are clear similarities between the measured traces and the simulated ones. The exception is the third set, where the graphs look very different. We assume that the peaks of the measured frequencies are the input frequencies and that compared to the simulation the spread of frequencies is not as good.
\\
\subsubsection{Multicore vs simulations}
Same as single core, but both simulated and tested using two cores. In~\autoref{fig:FFT_double} the measured frequencies are quite different from the simulated ones. 

\include{Figures/FFT_Images_single}

\include{Figures/FFT_Images_double}

\section{Analysis}
\label{sub:analysis}

The switching between two frequencies will occasionally create periods that are too short for the AES encryption, resulting in encryption error. Higher frequencies will cause this problem more frequently due to the increased amount of rising edges. Our simulations count the number of frequencies that are 4x faster than the base clock and use this as a measure of the risk of encryption errors. Removing these frequencies also gives a better view of the spectrum. 
\\
With the synchronization, the randomness from the frequencies is removed, and the attack results differ mainly due to how the frequencies affect the capture window and synchronization algorithm. We see that with slower frequencies, the capture window has to be expanded in order to capture the last round, increasing the memory requirement on the capture device in order to avoid a lower resolution. Having the resolution decreased will make it harder to capture the faster frequencies.
A dynamic range of frequencies will increase the quality needed of the capture device and also make it harder to write a synchronization algorithm.

When looking at the results in \autoref{tab:attack_results} we can use the difference between worst time overhead and the mean overhead to get an idea of how dynamic the frequencies are. Our hypothesis is that this difference should compare well with the number of removed traces. Notable here is that a high number of removed traces can have an effect on the worst timing overhead since some of the traces removed are due to failed synchronization, which can be caused if the capture window is too narrow for the time overhead. We believe that this will cause the worst overhead cases to be excluded and that the results from this column are unreliable. 

This is especially apparent in \hyperref[tab:attack_results]{"Two high, one above half, one lower than half"}\hyperref[tab:attack_results]{(4)}. Here the worst overhead is lower than the mean overhead, which should not be possible. This probably originates from the slower frequencies increasing the timing overhead, leading to failed synchronization and therefore being excluded from the used traces. The limitation of a capture window is not present in the simulation, and the simulated mean value is probably more reliable.
\\
Our original hypothesis was that frequencies lower than half of what the base clock is would lead to a much higher timing overhead. There appear to be cases where this is not entirely true, considering the mean overhead in \autoref{tab:attack_results} where some of the lower frequencies have a lower timing overhead. We believe this is because exception 3 - where the active frequency is low and switches to a high frequency - will create an extra period.
\\
In order to perform an attack on the system with a duplicate clock~\cite{Duplication}  the correct peak for the attack has to be guessed. Depending on how dynamically the frequencies appear, the possible permutations for the attack in order to find the correct peak can vary between \(2^n\) and \(11^n\) in a system of with 10 rounds, where \(n\) is the number of traces. In order to reach the higher value of permutations, the fastest frequencies have to be more than eleven times higher than the slowest. In order for this to happen some frequencies have to be so high that a lot of encryption errors will occur, and some frequencies will be very low and cause time overhead for the encryption. Simulations show that the bandwidth for our tested frequencies are around 20 - 25 MHz with the lower frequencies around the value 3-5 MHz. This indicates that without any extreme values for the fundamental frequencies the number of permutations for finding the correct peak is between  \(4^n\) and \(8^n\) most heavily dependent on how the lowest fundamental frequency is set.
\\

When both clocks overlap, i.e. when both AES cores have a rising edge at the same time, they will create a larger peak for the power trace that is easily identified.
 
\begin{figure}[H]
  \begin{subfigure}{0.5\textwidth}
    
    \includegraphics[width=\linewidth]{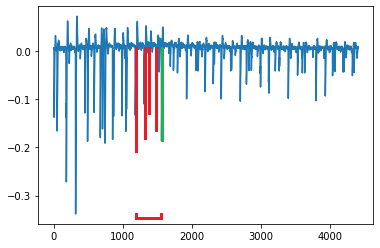}
   
    \caption{\centering Last round highlighted in green with the red peaks as potential correct peaks with a brute force of  \(5^n\) traces.}
  \end{subfigure}%
  \begin{subfigure}{0.5\textwidth}
    
    \includegraphics[width=\linewidth]{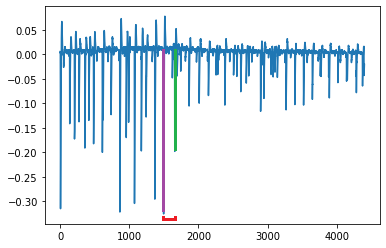}
    \caption{\centering Since the large peak(purple) correlates to both clocks it reduces the alternative for the correct last peak}
  \end{subfigure}
  \caption{\centering Captured traces from a duplicate clock with frequencies that would require a brute force of \(5^n\) traces. The green highlight shows where the last peak is. }
  \label{figure:Permutationslastpeak}
\end{figure}

This larger peak can be used to reduce the number of permutations in order to find the correct peak. For example: In a system where the correct peak can be guessed \(5^n\) a large peak next to the last peak will increase the probability of success from 20\% to 50\% for that trace. See \autoref{figure:Permutationslastpeak}

\section{Discussion}

When considering one clock our results and the \autoref{sub:analysis} work well and are accurate, but with two clocks in a system like~\cite{Duplication} some extra properties have to be taken into consideration. 

As mentioned in \autoref{sub:analysis}, some of the frequencies created will be faster than what the AES implementation can handle. Regarding the duplicated clock system used in~\cite{Duplication} it might be possible to use these encryption errors in a sort of reverse way. The two AES cores are clocked independently of each other and while both of the cores would have plenty of time to finish the encryption, the power trace measured in an attack might have peaks very close to each other. If these peaks are too close to each other, the rise and fall time could cause problems with the Hamming weight measures in an attack. This can also contribute to the fact that the simulations and testing differ a lot with the duplicated clock.

The two base clocks can be put differently relative to the frequencies in a way that they fulfil the exceptions discussed in \autoref{sec:methodintroduction} differently. While this creates for a more dynamic frequency spectrum, it might be possible to filter out the frequencies from the different base clocks if their spectra deviate too much from each other. To have both of the frequency spectra closely related but still unique, the same exceptions should be applied for both AES cores.

Attacking the first round with DL/MLP~\cite{Martin} but sorting out all traces that do not contain an overlap in the first round would make for a collection where the first peak in every trace correlates to the first round of AES with the possibility that the key can be retrieved. Simulations and testing shows that around 11\% of the traces for our values contains a larger peak for the first round. Since the last round only overlaps if the completion time for the previous rounds add up, it is highly unlikely for this method to work for a CPA.
\\

\section{Conclusions and Future Work}

\section{Conclusions}
\label{sec:conclusions}

Deciding what the optimal frequencies are for the setups used in~\cite{Martin} and the duplicated version~\cite{Duplication} is difficult. Depending on which system and what attack script is being used, different sets of frequencies seem to be optimal. It also seems to be a trade-off regarding how much time overhead is allowed. In our CPA attacks against the randomized single clock~\cite{Martin}  the results of \autoref{tab:attack_results} show that the optimal frequencies should be three frequencies above the base clock and one below half the value of the base clock. This resulted in few encryption errors, a high amount of traces removed and a low amount of time overhead. The worst case time overhead is not the best but this number is as stated in \autoref{sub:analysis} unreliable and the mean probably is a more relevant number in most cases.  If the time overhead is not a problem the one with two high frequencies and two below half the value of the base clock gives better result.

Based on the theory, simulation and testing in this thesis, the overlap happens mainly due to the two clocks switching to the same frequency setting two times in a row and creating a full synchronized period, frequencies faster than the base clock will occasionally make this happen in one base clock cycle too. Having all the fundamental frequencies set to below half the value of the base clock would increase the cumulative times both of the clocks have to pick the same frequency in order for this overlap to happen. This will camouflage the fundamental frequencies in the spectrum too, but then the base clock’s frequencies will appear more often. The base clocks have different frequencies, so they will not cause overlaps, but dominant frequencies might be bad in the sense that they might be filtered out. This strategy, as mentioned earlier, will cause a lot of time overhead and should only be used if the time overhead is a minor problem compared to security.

\section{Future work}
Due to time limitations and struggles during parts of the project, we were not able to meet all goals set. In this section, we will focus on some of the remaining issues that should be addressed in future work.

\subsection{What is not yet done?}
We have not been able to fully explore how the frequencies affect the safety in a dual AES core setup. Since someone is yet to find a way to break the implementation of~\cite{Duplication} using attacks as we did in \autoref{ch:DataValidity} was impossible. Exploring this and looking into ways of exploiting how frequencies affect the predictability of dual-core AES is yet to be done. As mentioned in the discussion section we have an idea of how to potentially attack the setup with DL/MPL. Due to time limits we were not able to study the concepts of DL/MPL in order to write an attack script for it.
\\
Throughout the thesis, we assume that the 4 fundamental frequencies all have a 50\% duty cycle. As described in exceptions 1 \& 3 the duty cycle affects the behaviour of the exceptions. This should be experimented more on in the future since the FPGA can produce different duty cycles for each of the 4 fundamental frequencies.


\printbibliography

\cleardoublepage
\end{document}

%% file: Figures/frequencyfigure.tex
\begin{figure}[H]
\centering
\begin{tikzpicture}

    \def\base{pi/5}
    \def\fone{5}
    \def\ftwo{7}
    \def\fthree{13}
    \def\ffour{21}
    
  \begin{axis}[
    width=\textwidth,
    height=10cm,
    xlabel={Time},
    samples=500,
    domain=0:2*pi,
    ymin=-12,
    ymax=8,
    axis lines=none,
    xtick={0,\base,2*\base,3*\base,4*\base, 5*\base, 6*\base, 7*\base, 8*\base, 9*\base, 10*\base},
    xticklabels={$0$, 1, 2,3, 4, 5, 6, 7, 8, 9, 10},
    ytick=\empty,
    clip=false
  ]
  
    \addplot+[red,line width=0.8pt,mark=none,domain=0.5:2*pi, solid] {5 + sign(sin(deg(5*x)))};
    
    \addplot+[blue,line width=0.8pt,mark=none,domain=0.5:2*pi, solid] {2 + sign(sin(deg(7*x)))};

    \addplot+[green,line width=0.8pt,mark=none,domain=0.5:2*pi, solid] {-1 + sign(sin(deg(13*x)))};

    \addplot+[orange,line width=0.8pt,mark=none,domain=0.5:2*pi, solid] {-4 + sign(sin(deg(21*x)))};

    \addplot+[black,line width=0.8pt,mark=none,domain=0.4:2*pi, solid] {-7 + sign(sin(deg(10*x)))};

     \addplot+[red,line width=0.8pt,mark=none,domain=0.5:\base, solid] {-10 + sign(sin(deg(\fone*x)))};
     \addplot+[red,line width=0.8pt,mark=none,domain=\base + 0.01:2*\base, solid] {-10 + sign(sin(deg(\fone*x)))};
     \addplot+[orange,line width=0.8pt,mark=none,domain=2*\base + 0.01:3* \base, solid] {-10 + sign(sin(deg(\ffour*x)))};
     \addplot+[blue,line width=0.8pt,mark=none,domain=3*\base + 0.01:4* \base, solid] {-10 + sign(sin(deg(\ftwo*x)))};
     \addplot+[blue,line width=0.8pt,mark=none,domain=4*\base + 0.01:5* \base, solid] {-10 + sign(sin(deg(\ftwo*x)))};
     \addplot+[green,line width=0.8pt,mark=none,domain=5*\base + 0.01:6* \base, solid] {-10 + sign(sin(deg(\fthree*x)))};
     \addplot+[green,line width=0.8pt,mark=none,domain=6*\base + 0.01:7* \base, solid] {-10 + sign(sin(deg(\fthree*x)))};
     \addplot+[red,line width=0.8pt,mark=none,domain=7*\base + 0.01:8* \base, solid] {-10 + sign(sin(deg(\fone*x)))};
     \addplot+[orange,line width=0.8pt,mark=none,domain=8*\base + 0.01:9* \base, solid] {-10 + sign(sin(deg(\ffour*x)))};
     \addplot+[red,line width=0.8pt,mark=none,domain=9*\base + 0.01:10* \base, solid] {-10 + sign(sin(deg(\fone*x)))};

    \pgfplotsinvokeforeach{1,2,...,10}{
        \addplot+[black,line width=0.8pt,mark=none,const plot, dashed] coordinates {(#1*pi/5, -12) (#1*pi/5,             7)};
            \node at (axis cs: #1*pi/5, -13) {$T_{#1}$};
     }

     \node at (axis cs: 0, 5) {F1:};
      \node at (axis cs: 0, 2) {F2:};
      \node at (axis cs: 0, -1) {F3:};
      \node at (axis cs: 0, -4) {F4:};
      \node at (axis cs: 0, -7) {Fbase:};
    \node at (axis cs: 0, -10) {Fout:};

  \end{axis}
  
\end{tikzpicture}
\caption{Four frequencies with base clock and output based on the setup used in \cite{Martin} . 
Exception 1 can be seen at $T_3$ where the current frequency does not end until between $T_4$ and $T_5$
Exception 2 lets the active frequency almost pass two cycles between $T_8$ and $T_9$. 
Exception 3 creates a rising edge at $T_2$. }

\label{figure:frequencies}
\end{figure}

%% file: Figures/Permutation_table.tex
\begin{table}[H]
\setlength{\arrayrulewidth}{0.5mm}
\setlength{\tabcolsep}{18pt}

\renewcommand{\arraystretch}{1.5}

\begin{tabular}{ |p{3cm}|p{3cm}|p{3cm}|  }

\hline
Number& Permutations &Total  \\
\hline
1 & \(4*(3+4)-4\) &24 \\
2 & \(4*(2+4)-8\)   & 16 \\
3 &\(4*(1+4)-12\) & 8 \\
4 &\(4*(0+4)-16\) & 0 \\

\hline
\end{tabular}
\caption{Permutations considering a number of frequencies without a rising edge}
\label{tab:Permutations}
\end{table}

%% file: Figures/Perm_without_rising_edge.tex
\begin{tikzpicture}[baseline=(M.center)]
    \matrix (M) [
        matrix of math nodes,
        nodes in empty cells,
        left delimiter={[},
        right delimiter={]},
        row sep=3em,
        column sep=3.7em,
    ]{
        F_{11} & \cancel{F_{12}} & F_{13}  \\
        F_{21} & F_{22} & F_{23} \\
        F_{31} & F_{32} & F_{33} \\
        F_{41} & F_{42} & F_{43} \\
    };

     \draw[red, dashed, ->] (M-1-1.east) -- (M-1-2.west);
    \draw[red, ->] (M-1-1.east) -- (M-2-2.west);
    \draw[red,->] (M-1-1.east) -- (M-3-2.west);
    \draw[red,->] (M-1-1.east) -- (M-4-2.west);
    
    \draw[red, dashed, bend left=10, ->] (M-1-2.east) to (M-1-3.west);
    \draw[blue, dashed, bend left=5, ->] (M-1-2.east) to (M-1-3.west);
    \draw[green, dashed, bend right=5, ->] (M-1-2.east) to (M-1-3.west);
    \draw[orange, dashed, bend right=10, ->] (M-1-2.east) to (M-1-3.west);

    \draw[red, dashed, bend left=10, ->] (M-1-2.east) to (M-2-3.west);
    \draw[blue, dashed, bend left=5, ->] (M-1-2.east) to (M-2-3.west);
    \draw[green, dashed, bend right=5, ->] (M-1-2.east) to (M-2-3.west);
    \draw[orange, dashed, bend right=10, ->] (M-1-2.east) to (M-2-3.west);

    \draw[red, dashed, bend left=10, ->] (M-1-2.east) to (M-3-3.west);
    \draw[blue, dashed, bend left=5, ->] (M-1-2.east) to (M-3-3.west);
    \draw[green, dashed, bend right=5, ->] (M-1-2.east) to (M-3-3.west);
    \draw[orange, dashed, bend right=10, ->] (M-1-2.east) to (M-3-3.west);

    \draw[red, dashed, bend left=10, ->] (M-1-2.east) to (M-4-3.west);
    \draw[blue, dashed, bend left=5, ->] (M-1-2.east) to (M-4-3.west);
    \draw[green, dashed, bend right=5, ->] (M-1-2.east) to (M-4-3.west);
    \draw[orange, dashed, bend right=10, ->] (M-1-2.east) to (M-4-3.west);

    \draw[blue, dashed, ->] (M-2-1.east) to (M-1-2.west);
    \draw[blue, ->] (M-2-1.east) -- (M-2-2.west);
    \draw[blue, ->] (M-2-1.east) -- (M-3-2.west);
    \draw[blue, ->] (M-2-1.east) -- (M-4-2.west);

    \draw[green, dashed, ->] (M-3-1.east) to (M-1-2.west);
    \draw[green,->] (M-3-1.east) -- (M-2-2.west);
    \draw[green,->] (M-3-1.east) -- (M-3-2.west);
    \draw[green,->] (M-3-1.east) -- (M-4-2.west);

    \draw[orange, dashed, ->] (M-4-1.east) to (M-1-2.west);
    \draw[orange,->] (M-4-1.east) -- (M-2-2.west);
    \draw[orange,->] (M-4-1.east) -- (M-3-2.west);
    \draw[orange,->] (M-4-1.east) -- (M-4-2.west);

\end{tikzpicture}

%% file: Figures/Perm_after_rising_edge.tex
\begin{tikzpicture}[baseline=(M.center)]
    \matrix (M) [
        matrix of math nodes,
        nodes in empty cells,
        left delimiter={[},
        right delimiter={]},
        row sep=3em,
        column sep=3.7em,
    ]{
        F_{11} & \cancel{F_{12}} & F_{13}  \\
        F_{21} & F_{22} & F_{23} \\
        F_{31} & F_{32} & F_{33} \\
        F_{41} & F_{42} & F_{43} \\
    };

     \draw[red, dashed, ->] (M-1-1.east) -- (M-1-2.west);
    \draw[red, dashed,  ->] (M-1-2.east) -- (M-1-3.west);
    \draw[red, dashed,  ->] (M-1-2.east) -- (M-2-3.west);
    \draw[red, dashed, ->] (M-1-2.east) -- (M-3-3.west);
    \draw[red, dashed, ->] (M-1-2.east) -- (M-4-3.west);

    \draw[blue, dashed, ->] (M-2-1.east) to (M-1-2.west);
     \draw[blue, ->] (M-2-2.east) -- (M-1-3.west);
    \draw[blue, ->] (M-2-2.east) -- (M-2-3.west);
    \draw[blue, ->] (M-2-2.east) -- (M-3-3.west);
    \draw[blue, ->] (M-2-2.east) -- (M-4-3.west);

    \draw[green, dashed, ->] (M-3-1.east) to (M-1-2.west);
    \draw[green,->] (M-3-2.east) -- (M-1-3.west);
    \draw[green,->] (M-3-2.east) -- (M-2-3.west);
    \draw[green,->] (M-3-2.east) -- (M-3-3.west);
    \draw[green,->] (M-3-2.east) -- (M-4-3.west);

    \draw[orange, dashed, ->] (M-4-1.east) to (M-1-2.west);
    \draw[orange,->] (M-4-2.east) -- (M-1-3.west);
    \draw[orange,->] (M-4-2.east) -- (M-2-3.west);
    \draw[orange,->] (M-4-2.east) -- (M-3-3.west);
    \draw[orange,->] (M-4-2.east) -- (M-4-3.west);

\end{tikzpicture}

%% file: Figures/Probability_no_rising_edge.tex
\begin{equation}
\begin{aligned}
P(\text{Zero rising edges}) &= (1-p_1) \cdot (1-p_2) \cdot (1-p_3) \cdot (1-p_4) \\
\\
P(\text{One rising edge}) &= p_1 \cdot (1-p_2) \cdot (1-p_3) \cdot (1-p_4) \\
&\quad + p_2 \cdot (1-p_1) \cdot (1-p_3) \cdot (1-p_4) \\
&\quad + p_3 \cdot (1-p_1) \cdot (1-p_2) \cdot (1-p_4) \\
&\quad + p_4 \cdot (1-p_1) \cdot (1-p_2) \cdot (1-p_3) \\
\\
P(\text{Two rising edges}) &= p_1 \cdot p_2 \cdot (1-p_3) \cdot (1-p_4) \\
&\quad + p_1 \cdot p_3 \cdot (1-p_2) \cdot (1-p_4) \\
&\quad + p_1 \cdot p_4 \cdot (1-p_2) \cdot (1-p_3) \\
&\quad + p_2 \cdot p_3 \cdot (1-p_1) \cdot (1-p_4) \\
&\quad + p_2 \cdot p_4 \cdot (1-p_1) \cdot (1-p_3) \\
&\quad + p_3 \cdot p_4 \cdot (1-p_1) \cdot (1-p_2) \\
\\
P(\text{Three rising edges}) &= p_2 \cdot p_3 \cdot p_4 \cdot (1 - p_1) \\
&\quad + p_1 \cdot p_3 \cdot p_4 \cdot (1 - p_2) \\
&\quad + p_1 \cdot p_2 \cdot p_4 \cdot (1 - p_3) \\
&\quad + p_1 \cdot p_2 \cdot p_3 \cdot (1 - p_4) \\
\\
P(\text{Four rising edges}) &= p_1 \cdot p_2 \cdot p_3 \cdot p_4 \\
\end{aligned}
\label{equation:perm_with_certain_number_of_RE}
\end{equation}

%% file: Figures/newSim.tex
\begin{longtable}{| p{1.6cm} | >{\scriptsize \centering}p{1.6cm}|p{6cm}|  p{1.2cm} | p{1cm}|}

\hline
\multicolumn{5}{|c|}{Simulated frequencies} \\
\hline
Frequency group &  {\footnotesize Frequencies (MHz)} & Graph &    Nr. freqs &  Unique freqs\\
\hline
Previous work &  $f_{base}$=10 $f_1$=11.9713 $f_2$=7.7315 $f_3$=9.2778  $f_4$=12.6515 & \raisebox{-\totalheight} {\includegraphics[width=60mm, height=30mm]{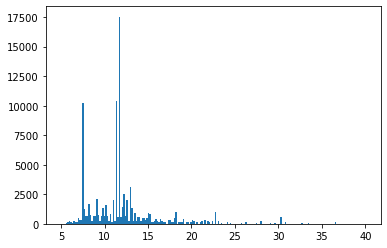}} & 38872 &  412  \\
\hline
Two low, one above half, one lower than half &   $f_{base}$=10 $f_1$=9.5917 $f_2$=9.0317 $f_3$=6.2777 $f_4$=4.0517 & \raisebox{-\totalheight} {\includegraphics[width=60mm, height=30mm]{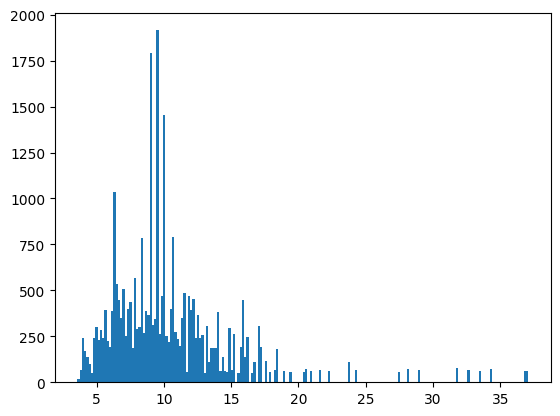}} & 28610 & 458 \\
\hline
Two high, two lower than half &   $f_{base}$=10 $f_1$=3.6719 $f_2$=4.4021 $f_3$=12.9781 $f_4$=14.4317 & \raisebox{-\totalheight} {\includegraphics[width=60mm, height=30mm]{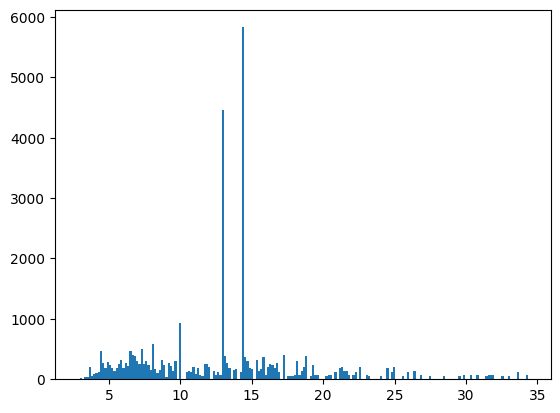}} & 33992 & 502\\
\hline
Three high, one lower than half &   $f_{base}$=10 $f_1$=4.7717 $f_2$=11.5019 $f_3$=12.0779 $f_4$=13.5319 & \raisebox{-\totalheight} {\includegraphics[width=60mm, height=30mm]{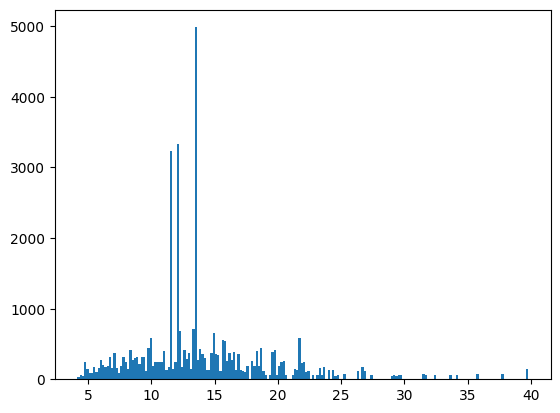}} & 38910 & 496\\
\hline
Three low, one lower than half &   $f_{base}$=10 $f_1$=9.2003 $f_2$=9.3001 $f_3$=9.4001 $f_4$=4.4003 & \raisebox{-\totalheight} {\includegraphics[width=60mm, height=30mm]{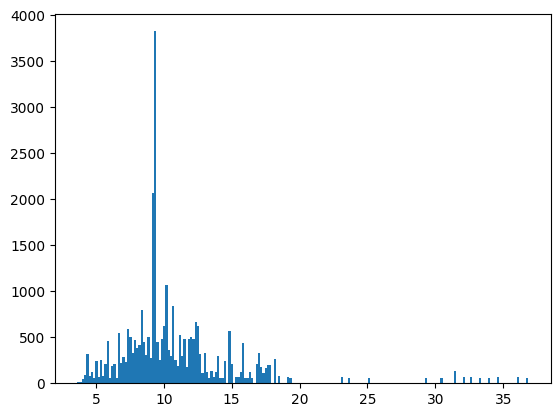}} & 30796 & 458\\
\hline
Three high, one above half &  $f_{base}$=10 $f_1$=5.9009 $f_2$=11.5019 $f_3$=12.0779 $f_4$=13.5319 & \raisebox{-\totalheight} {\includegraphics[width=60mm, height=30mm]{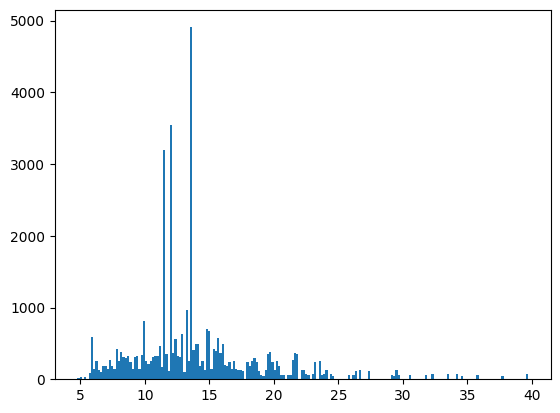}} & 39929 & 468\\
\hline
Two high, one above half, one lower than half &  $f_{base}$=10 $f_1$=11.8713 $f_2$=10.6017 $f_3$=5.1779 $f_4$=3.6317 & \raisebox{-\totalheight} {\includegraphics[width=60mm, height=30mm]{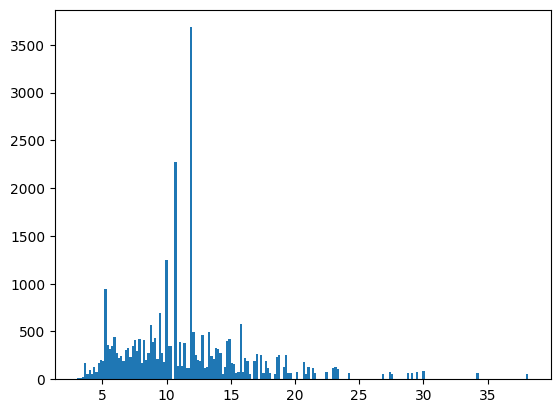}} & 91369 & 504\\
\hline

\caption{Frequency test on hardware}

\label{tab:simulation_results}
\end{longtable}
\begin{center}
 \vspace*{-5mm}

{ \footnotesize  The frequencies generated by all sets of input frequencies are displayed in this table. Each row contains one set of frequencies and has the total number of frequencies generated as well as the number of unique frequencies. The first column represents the relation between the frequencies and the base frequency.}
\end{center}

%% file: Figures/Trace_table.tex
\begin{table}[H]
\centering
{\footnotesize
\begin{tabular}{|>{\centering}p{0.2cm}|>{\footnotesize}p{2.2cm}|>{\centering} p{1.2cm}| 
>{\centering} p{1.2cm}| >{\centering} p{1.2cm}|  >{\centering} p{1cm}|  >{\centering} p{1.2cm} | p{1.1cm} |}
\hline
\multicolumn{8}{|c|}{Frequencies used and traces required to break} \\
\hline
 & {\footnotesize Frequencies} & Nr. Traces & Failed Enc & Removed Traces & Max delay & Worst overhead & Mean overhead\\
\hline
1 & \hyperref[tab:simulation_results]{Previous
work}& \textcolor{red}{3750} & 0.8\% & 20\% & 2300 & 30\% & -18\% \\
\hline
2 & \hyperref[tab:simulation_results]{Two low, one above half, one lower than half} & 4750 & 5.9\% & \textcolor{green}{47\%} & 2040 & 15\% & \textcolor{red}{12\%}  \\
\hline
3 & \hyperref[tab:simulation_results]{Two high, two lower than half} & \textcolor{green}{5500} & \textcolor{green}{0\%} &  31\% & \textcolor{red}{3800} & \textcolor{red}{125\%} & -6\%\\
\hline
4 &\hyperref[tab:simulation_results]{Two high, one above half, one lower than half} & 4500 & \textcolor{red}{13\%} & \textcolor{red}{17\%} & \textcolor{green}{1800} & \textcolor{green}{0\%} & 5\%\\
\hline
5 &\hyperref[tab:simulation_results]{Three high, one lower than half}& 4750 & 0.11\% & 40.5\% & 2680 & 50\% & \textcolor{green}{-20\%}\\
\hline
6 &\hyperref[tab:simulation_results]{Three low, one lower than half} & 4750 & 0.07\% & 41\% & 2000 & 12\% & 4\%\\
\hline
7 & \hyperref[tab:simulation_results]{Three high, one above half} & 4000 & 3.6\% & 27\% & 2250 & 26\% & -18\%\\
\hline

\end{tabular}
}
\caption{Frequency test on hardware}
{\footnotesize Describes the different properties of frequency combinations. Col 1: Relations of frequencies used, detailed in \autoref{tab:simulation_results} Col 2: Number of traces required to get the encryption key. Col 3: Percentage of failed encryptions where the returned ciphertext was invalid. Col 4: Percentage of removed traces, where synchronization and noise filtration removed traces that were invalid or hard to read. Col 5: Worst case timestamp of the peak from the last round. Col 6: Gives the percentage of the column to the left compared to the base frequency. Col 7: The simulated timing mean overhead in relation to a none randomized clock.
}
\label{tab:attack_results}
\end{table}

%% file: Figures/Relation_comparison.tex
\begin{table}[H]
\centering
{\footnotesize
\begin{tabular}{|>{\scriptsize \centering}p{2.2cm}|>{\centering} p{1.2cm}| 
>{\centering} p{1.2cm}| >{\centering} p{1.2cm}|  >{\centering} p{1.2cm}|  >{\centering} p{1.2cm} | p{1cm} |}
\hline
\multicolumn{7}{|c|}{ Frequencies used and traces required to break} \\
\hline
{\footnotesize Frequencies} & Nr. Traces & Failed Enc & Removed Traces & Max delay & Timing overhead & Sim. overhead\\
\hline
\multicolumn{7}{|c|}{\hyperref[tab:simulation_results]{Two high, two lower than half}} \\											
\hline
$f_{base}$=10 $f_1$=14.4317 $f_2$=12.9781 $f_3$=4.4021  $f_4$=3.6719 & 5500 & 0\% & 31\% & 3800 & 125\% & -7\%  \\
\hline
$f_{base}$=10 $f_1$=14.5370 $f_2$=13.0883 $f_3$=4.3611  $f_4$=3.6725 & 5000 & 0.02\% & 30\% & 4200 & 136\% & -6\%  \\
\hline
$f_{base}$=10 $f_1$=11.8182 $f_2$=10.8333 $f_3$=4.8148  $f_4$=3.5017 & 6000 & 6.3\% & 34\% & 4300 & 140\% & 6\%  \\

\hline
\multicolumn{7}{|c|}{\hyperref[tab:simulation_results]{Three low, one lower than half}} \\
\hline
$f_{base}$=10 $f_1$=9.7991 $f_2$=9.1458 $f_3$=8.9959  $f_4$=4.4254 & 4500 & 0.52\% & 32\% & 2030 & 14\% & 2\%  \\
\hline
$f_{base}$=10 $f_1$=9.4001 $f_2$=9.3001 $f_3$=9.2003  $f_4$=4.003 & 4750 & 0.07\% & 41\% & 2000 & 12\% & 6\%  \\
\hline
$f_{base}$=10 $f_1$=9.8659 $f_2$=9.4485 $f_3$=8.9527  $f_4$=4.5009 & 4500 & 0.034\% & 23\% &2800 & 23\% & 10\%  \\
\hline
\end{tabular}
}
\caption{Frequency test on hardware}
{\footnotesize Contains the same information as \autoref{tab:attack_results} but instead of comparing the relation between frequencies and the base clock it compares specific frequencies. Proving whether small changes in frequencies have a effect on the results. Two types of frequencies are tested, with three sets of frequencies for each used. 
}
\label{tab:relation_comparison}
\end{table}

%% file: Figures/FFT_Images_single.tex
\begin{figure}[t!]
\centering
\captionsetup[subfigure]{labelformat=empty}

  \begin{subfigure}[b]{0.44\linewidth}
    \includegraphics[width=64mm, height=33mm]{Images/Single/Default.png}
    \vspace*{-8mm}
    \caption{\scriptsize FFT of \hyperref[tab:simulation_results]{Previous
work}}
  \end{subfigure}
   \hspace{7mm}
    \begin{subfigure}[b]{0.44\linewidth}
    \includegraphics[width=65mm, height=33mm]{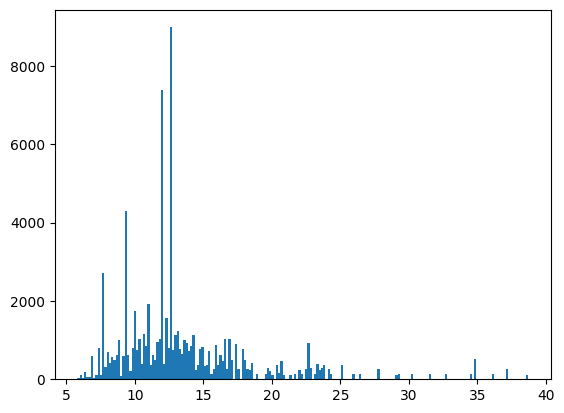}
    \vspace*{-8mm}
    \caption{\scriptsize \centering Sim of \hyperref[tab:simulation_results]{Previous
work}}
  \end{subfigure}

   \begin{subfigure}[b]{0.45\linewidth}
    \includegraphics[width=65mm, height=33mm]{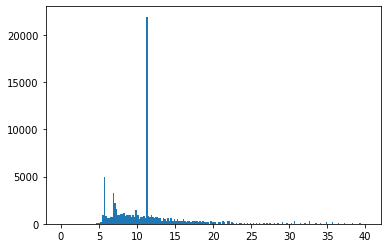}
    \vspace*{-8mm}
    \caption{\scriptsize FFT of \hyperref[tab:simulation_results]{Three high, one lower than half}}
  \end{subfigure}
 \hspace{7mm}
\begin{subfigure}[b]{0.45\linewidth}
    \includegraphics[width=65mm, height=33mm]{Images/Single/cl10.0,_5.9009,_11.5019,_12.0779,_13.5319.png}
    \vspace*{-8mm}
    \caption{\scriptsize Sim of \hyperref[tab:simulation_results]{Three high, one lower than half}}
  \end{subfigure}

   \begin{subfigure}[b]{0.45\linewidth}
    \includegraphics[width=65mm, height=33mm]{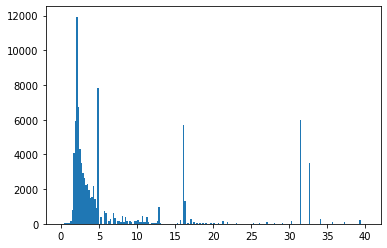}
    \vspace*{-8mm}
    \caption{\scriptsize FFT of \hyperref[tab:simulation_results]{Two high, two lower than half}}
  \end{subfigure}
    \hspace{7mm}
      \begin{subfigure}[b]{0.45\linewidth}
    \includegraphics[width=65mm, height=33mm]{Images/Single/cl10.0,_3.6719,_4.4021,_12.9781,_14.4317.png}
    \vspace*{-8mm}
    \caption{\scriptsize Sim of \hyperref[tab:simulation_results]{Two high, two lower than half}}
  \end{subfigure}

   \begin{subfigure}[b]{0.45\linewidth}
    \includegraphics[width=65mm, height=33mm]{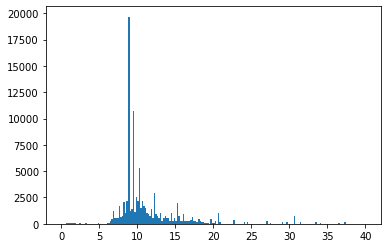}
    \vspace*{-8mm}
    \caption{\scriptsize FFT of \hyperref[tab:simulation_results]{Two low, one above half, one lower than half}}
  \end{subfigure}
   \hspace{7mm}
   \begin{subfigure}[b]{0.45\linewidth}
    \includegraphics[width=65mm, height=33mm]{Images/Single/cl10,_9.5917,_9.0317,_6.2777,_4.0517.png}
    \vspace*{-8mm}
    \caption{\scriptsize Sim of \hyperref[tab:simulation_results]{Two low, one above half, one lower than half}}
  \end{subfigure}

   \begin{subfigure}[b]{0.45\linewidth}
    \includegraphics[width=65mm, height=33mm]{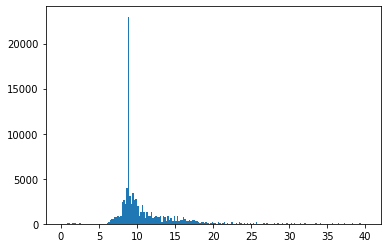}
    \vspace*{-8mm}
    \caption{\scriptsize FFT of \hyperref[tab:simulation_results]{Three low, one lower than half}}
  \end{subfigure}
   \hspace{7mm}
   \begin{subfigure}[b]{0.45\linewidth}
    \includegraphics[width=65mm, height=33mm]{Images/Single/cl10.0,_9.2003,_9.3001,_9.4001,_4.4003.png}
    \vspace*{-8mm}
    \caption{\scriptsize Sim of \hyperref[tab:simulation_results]{Three low, one lower than half}}
  \end{subfigure}
\caption{Fast Fourier Transform Single Core}
{\scriptsize Fast Fourier Transform of the measured power traces compared to the simulated frequencies of the same set of input frequencies. The figures show the spread of actual frequencies measured during testing (continued).}
\label{fig:FFT_single}
\end{figure}

%% file: Figures/FFT_Images_double.tex
\begin{figure}[!t]
\centering
\captionsetup[subfigure]{labelformat=empty}

  \centering
  \begin{subfigure}[b]{0.44\linewidth}
    \includegraphics[width=65mm, height=33mm]{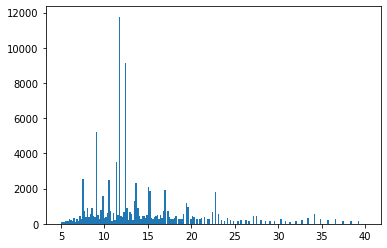}
    \vspace*{-8mm}
    \caption{\scriptsize FFT of \hyperref[tab:simulation_results]{Previous
work}}
  \end{subfigure}
    \hspace{7mm}
    \begin{subfigure}[b]{0.44\linewidth}
    \includegraphics[width=65mm, height=33mm]{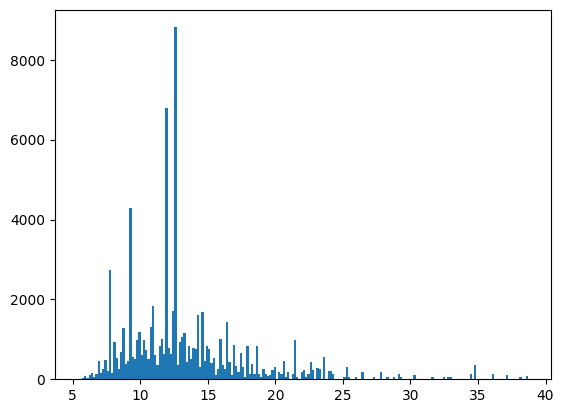}
    \vspace*{-8mm}
    \caption{\scriptsize Sim of \hyperref[tab:simulation_results]{Previous
work}}
  \end{subfigure}

   \begin{subfigure}[b]{0.45\linewidth}
    \includegraphics[width=65mm, height=33mm]{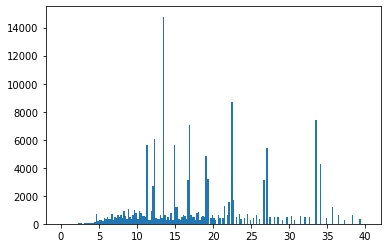}
    \vspace*{-8mm}
    \caption{\scriptsize FFT of \hyperref[tab:simulation_results]{Three high, one lower than half}}
  \end{subfigure}
  \hspace{7mm}
\begin{subfigure}[b]{0.45\linewidth}
    \includegraphics[width=65mm, height=33mm]{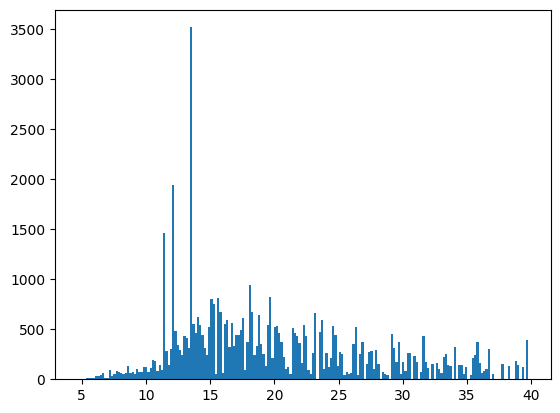}
    \vspace*{-8mm}
    \caption{\scriptsize Sim of \hyperref[tab:simulation_results]{Three high, one lower than half}}
  \end{subfigure}

   \begin{subfigure}[b]{0.45\linewidth}
    \includegraphics[width=65mm, height=33mm]{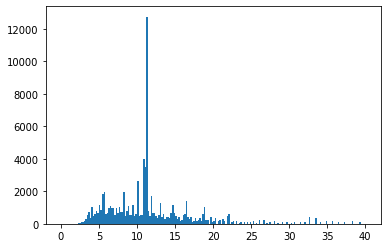}
    \vspace*{-8mm}
    \caption{\scriptsize FFT of \hyperref[tab:simulation_results]{Two high, two lower than half}}
  \end{subfigure}
  \hspace{7mm}
      \begin{subfigure}[b]{0.45\linewidth}
    \includegraphics[width=65mm, height=33mm]{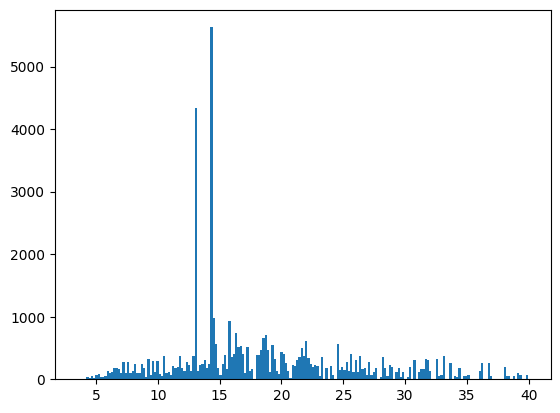}
    \vspace*{-8mm}
    \caption{\scriptsize Sim of \hyperref[tab:simulation_results]{Two high, two lower than half}}
  \end{subfigure}

   \begin{subfigure}[b]{0.45\linewidth}
    \includegraphics[width=65mm, height=33mm]{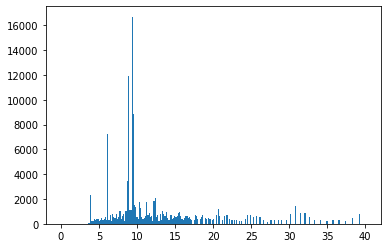}
    \vspace*{-8mm}
    \caption{\scriptsize FFT of \hyperref[tab:simulation_results]{Two low, one above half, one lower than half}}
  \end{subfigure}
    \hspace{7mm}
   \begin{subfigure}[b]{0.45\linewidth}
    \includegraphics[width=65mm, height=33mm]{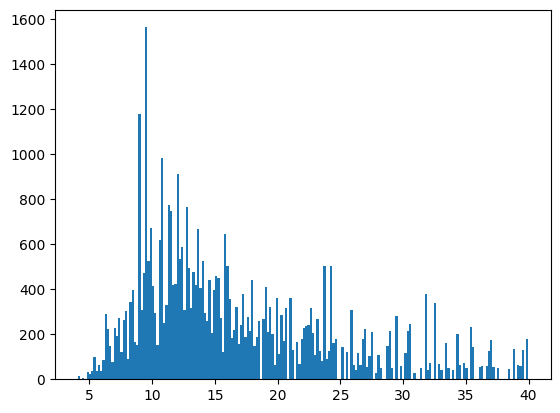}
    \vspace*{-8mm}
    \caption{\scriptsize Sim of \hyperref[tab:simulation_results]{Two low, one above half, one lower than half}}
  \end{subfigure}

   \begin{subfigure}[b]{0.45\linewidth}
    \includegraphics[width=65mm, height=33mm]{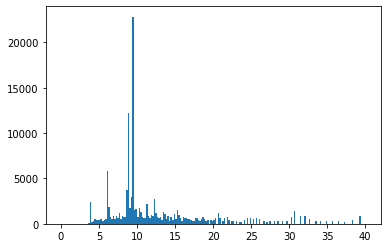}
    \vspace*{-8mm}
    \caption{\scriptsize FFT of \hyperref[tab:simulation_results]{Three low, one lower than half}}
  \end{subfigure}
    \hspace{7mm}
   \begin{subfigure}[b]{0.45\linewidth}
    \includegraphics[width=65mm, height=33mm]{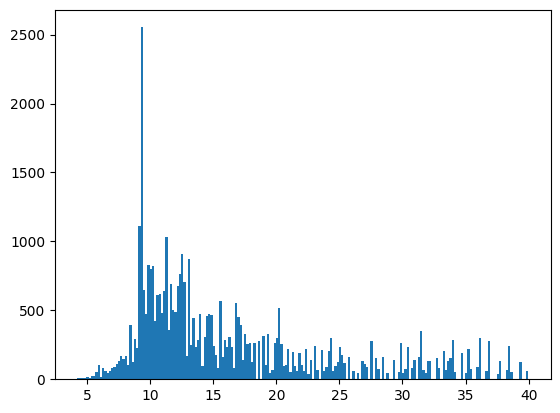}
    \vspace*{-8mm}
    \caption{\scriptsize Sim of \hyperref[tab:simulation_results]{Three low, one lower than half}}
  \end{subfigure}

  \caption {Fast Fourier Transform Dual Core}
  {\footnotesize Fast Fourier Transform of the measured power traces of a duplicated core compared to the simulated frequencies of the same set of input frequencies. The figures show the spread of actual frequencies measured during testing. }
  \label{fig:FFT_double}
\end{figure}